\documentclass[aps,pra,epsf,superscriptaddress,amsmath,amssymb,amsfonts,twocolumn,showpacs,nofootinbib,10pt]{revtex4-1}

\usepackage{graphicx}
\usepackage{epsfig}
\usepackage{dcolumn}
\usepackage{bm}
\usepackage{braket}
\usepackage{amsmath}
\usepackage{mathtools}
\usepackage{graphicx,color,xcolor}
\usepackage{hyperref}
\newcommand{\abs}[1]{\left| #1 \right|} 
\usepackage{hyperref}
\usepackage{multirow}

\allowdisplaybreaks

\usepackage[normalem]{ulem} 

\begin{document}

\title{Generation of dipolar supersolids through a  barrier sweep in droplet lattices}

\author{E. L. Brakensiek}
\affiliation{Department of Physics and LAMOR, Missouri University of Science and Technology, Rolla, MO 65409, USA}
\author{G. A. Bougas}
\affiliation{Department of Physics and LAMOR, Missouri University of Science and Technology, Rolla, MO 65409, USA} 
\author{S. I. Mistakidis}
\affiliation{Department of Physics and LAMOR, Missouri University of Science and Technology, Rolla, MO 65409, USA}

\date{\today}

\begin{abstract} 
 
We propose a dynamical protocol to generate supersolids in dipolar quantum gases by sweeping a repulsive Gaussian barrier through an incoherent quasi-one-dimensional droplet array.
Supersolidity is inferred by monitoring the ensuing dynamics of the density, momentum distribution, center-of-mass motion, and superfluid fraction within the framework of the extended Gross-Pitaevskii equation with quantum corrections.
A persistent superfluid background arises, atop which the crystals oscillate in unison, indicating the establishment of phase coherence. 
This process is accompanied by energy redistribution and the gradual transfer of higher-lying momenta toward the zero momentum mode. 
The dependence of the superfluid fraction on the barrier velocity and height is also elucidated evincing the parametric regions which facilitate the rise of a superfluid background. 
Our results pave the way for engineering supersolid generation using experimentally accessible protocols.

\end{abstract}

\maketitle

\section{Introduction}

Ultracold dipolar quantum gases offer pristine platforms for unraveling the effect of long-range anisotropic interactions which give rise to enriched many-body phases of matter and complex nonequilibrium behavior~\cite{bottcher2021new,chomaz_dipolar_2022,mukherjee2023droplets}. 
These interactions originate from either the magnetic moments of the atoms~\cite{chomaz_dipolar_2022} or the electric dipole moments of polar molecules~\cite{bigagli2024observation,schindewolf2025few}. 
Utilizing magnetic atoms, which we focus herein, dipolar Bose-Einstein condensates (dBECs) have been experimentally realized first with Chromium atoms~\cite{griesmaier2005bose}, and subsequently with lanthanides such as Dysprosium~\cite{lu2011strongly}, Erbium~\cite{aikawa2012bose} and Europium~\cite{miyazawa2022bose}. 
Here, the interplay between long- and short-range interactions gives rise to the softening of the roton mode~\cite{Lasinio_roton_2013,Bisset_roton_2013,jona2013roton,petter_probing_2019,Schmidt_Roton_excitation2021} facilitating wave collapse that is prevented in the presence of quantum fluctuations commonly modeled by the Lee-Huang-Yang (LHY) correction term~\cite{lee1957eigenvalues}. 
This mechanism gives rise to exotic states dubbed supersolids~\cite{Tanzi_observation_2019,chomaz2019long} and droplets~\cite{bottcher2019transient} that have been observed in the laboratory, but also more complex configurations~\cite{Hertkorn_pattern_2021,poli_maintaining_2021,zhang2019supersolidity,ripley_two-dimensional_2023} such as pumpkin, honeycomb, and labyrinthines that are yet to be realized. 
A standard theory framework describing adequately the aforementioned magnetic states and incorporating the LHY contribution is the extended Gross-Pitaevskii model~\cite{Lima_quantum_2011,santos2016fluctuations}, see also Refs~\cite{Buchler_MC,Bombin_MC,Kurbakov_MC,fedorov2014two,Macia_MC} for advanced ab-initio methods.  This framework is much less accurate for polar molecules due to the excessively larger dipolar interactions~\cite{cardinale2025exploring}.

Supersolids are among the most intriguing dipolar phase of matter as they combine the frictionless flow of a superfluid with the periodic density modulation of a crystal~\cite{recati_supersolidity_2023}. 
Hence, they simultaneously break the ${\rm U}(1)$ gauge symmetry associated with global phase coherence and the continuous translational symmetry yielding crystalline order~\cite{bottcher2019transient,Tanzi2019b,chomaz2019long}. 
Due to the spontaneously broken translational symmetry, these states accommodate additional sound mode branches~\cite{Blakie_sound_modes,Bougas_signatures_2025,cook2026excitations,Hertkorn_sound}. 
Besides magnetic atoms, supersolidity has been detected in completely different settings ranging from Rydberg gases~\cite{Henkel_Rydberg,Cinti_Rydberg}, spin-orbit coupled condensates~\cite{li2017stripe}, atoms in cavity fields~\cite{leonard2017monitoring,leonard2017supersolid} and more recently in the dynamics of parametrically driven short-range bosonic systems~\cite{liebster2025supersolid}. 
The supersolid properties have been intensively studied in the past few years aiming to understand the solid nature of the supersolid phase. 
Notable insights have been gained, for instance, from the excitation spectrum of these states via Bogoliubov de-Gennes analysis~\cite{Ronen_BdG_2006,hertkorn_supersolidity_2021,cook2026excitations}, their collective excitations~\cite{ferrier-barbut_scissors_2018,roccuzzo2022moment,zhen2025breaking}, the impact of the external trap~\cite{poli_maintaining_2021,li_merging_2023} dictating the symmetry of the ensuing structures, the characteristics of quantum vortices~\cite{Recati_vortex,casotti_observation_2024,Das_vortex}, associated tunneling processes~\cite{Mistakidis2024,Biagioni_measurement_2024}, the emergence of shear waves~\cite{Senarath_shear} and their elastic properties~\cite{Platt_sounds_2024,poli_excitations_2024,rakic2024elastic}.

Despite the realization of supersolids in dipolar quantum gases via interaction quenches~\cite{Chomaz2018,Tanzi_observation_2019,Halder_control_2022}, an open question remains regarding the mechanisms that catalyze the dynamical generation of supersolidity.
Recent work revealed that supersolid features are facilitated when a dipolar gas is heated, leading to a counterintuitive roton softening~\cite{sanchez_heating_2023}.
Moreover, it was demonstrated that the presence of a minority component in binary dipolar mixtures favors the creation of supersolids in an otherwise unmodulated phase~\cite{Scheiermann_catalyzation_2023}.
In a similar vein, the introduction of fermionic impurities in a bosonic dipolar gas also promotes supersolidity~\cite{Lewkowicz_dynamic_2025,Lewkowicz_supersolidity_2025}.

Here, we take an alternative route and
propose the dynamical generation of a supersolid starting from a fully incoherent elongated droplet lattice which is externally perturbed by a repulsive Gaussian barrier. 
The latter facilitates the inelastic collision of the initially incoherent crystals resulting in the gradual build-up of a superfluid background. 
Our three dimensional dipolar gas consists of Dysprosium atoms and it features weak (tight) harmonic confinement along the axial (transverse) direction. 
It is initiated in its ground-state quasi-one-dimensional (quasi-1D) droplet lattice structure in the suitable interaction regime, and it is subsequently perturbed by the barrier, see Fig.~\ref{fig:schematic}. 
Recall that similar barrier sweeps have been widely used in short-range superfluids to create solitary wave structures~\cite{Hakim_soliton,Brazhnyi_DS,Engels_obstacle,Katsimiga_obstacle}, dispersive shock waves~\cite{El_DSW,Kamchatnov_DSW,Hoefer_DSW} and vortical patterns~\cite{Onofrio_obstacle,carretero2007soliton,lim2022vortex}.

We find that the potential barrier sweeps induce inelastic collisions accompanied by particle tunneling between the originally incoherent droplet crystals. 
This process gives rise to a superfluid background coexisting with the crystalline density modulations which is a hallmark of supersolidity~\cite{Tanzi2019b,recati_supersolidity_2023}. 
Notably, this behavior occurs within the droplet interaction regime, i.e. the relative strength between dipolar and short-range interactions is larger than unity.

The dynamical formation of supersolidity is testified through characteristic measures.
In particular, following the build-up of a superfluid background the post-collision droplet crystals feature synchronized oscillations upon the finite background. 
This is suggestive of the establishment of global phase coherence, while the persistence of the crystal structures provides evidence of their rigid character~\cite{Bougas_signatures_2025}. 
The frequency of the in-sync crystal oscillations is significantly low compared to the one of the trap, which is a characteristic of collective excitations of dipolar supersolids~\cite{Hertkorn_decoupled_2024,Guo2019}.
Moreover, a gradual transfer from higher momenta toward lower-lying ones and notably to a pronounced zero-momentum peak takes place after the barrier drag, which is another manifestation of supersolidity~\cite{recati_supersolidity_2023}.
The latter is also evinced by the increase of the superfluid fraction following the barrier drag.
The superfluid fraction serves as an order parameter, delineating the supersolid formation regimes with respect to both the velocity and height of the barrier. For instance, at moderate velocities the superfluid background is pronounced, while after a critical velocity the droplet array remains almost un-perturbed and the superfluid fraction is suppressed.

This work is structured as follows. 
Section~\ref{sec:theory} describes the considered dipolar Dysprosium gas along with the suitable extended Gross-Pitaevskii framework and the sweep protocol of the external Gaussian barrier. 
In Section~\ref{sec:dynamics}, we elaborate on the dynamical nucleation of supersolid configurations by employing suitable diagnostic measures. 
We conclude and discuss future research directions inspired by our results in Section~\ref{sec:conclusions}. 
Appendix~\ref{app:3body} showcases the persistence of supersolid features
in the presence of three-body recombination processes. 
In Appendix~\ref{app:different_crystals} 
we address the impact of the size of the initial droplet array on the supersolid creation, while in Appendix~\ref{app:simulations} further details are provided on our numerical simulations.

\section{Droplet array and sweep protocol}\label{sec:theory}

\begin{figure}[t!]
\centering\includegraphics[width=0.95\linewidth]{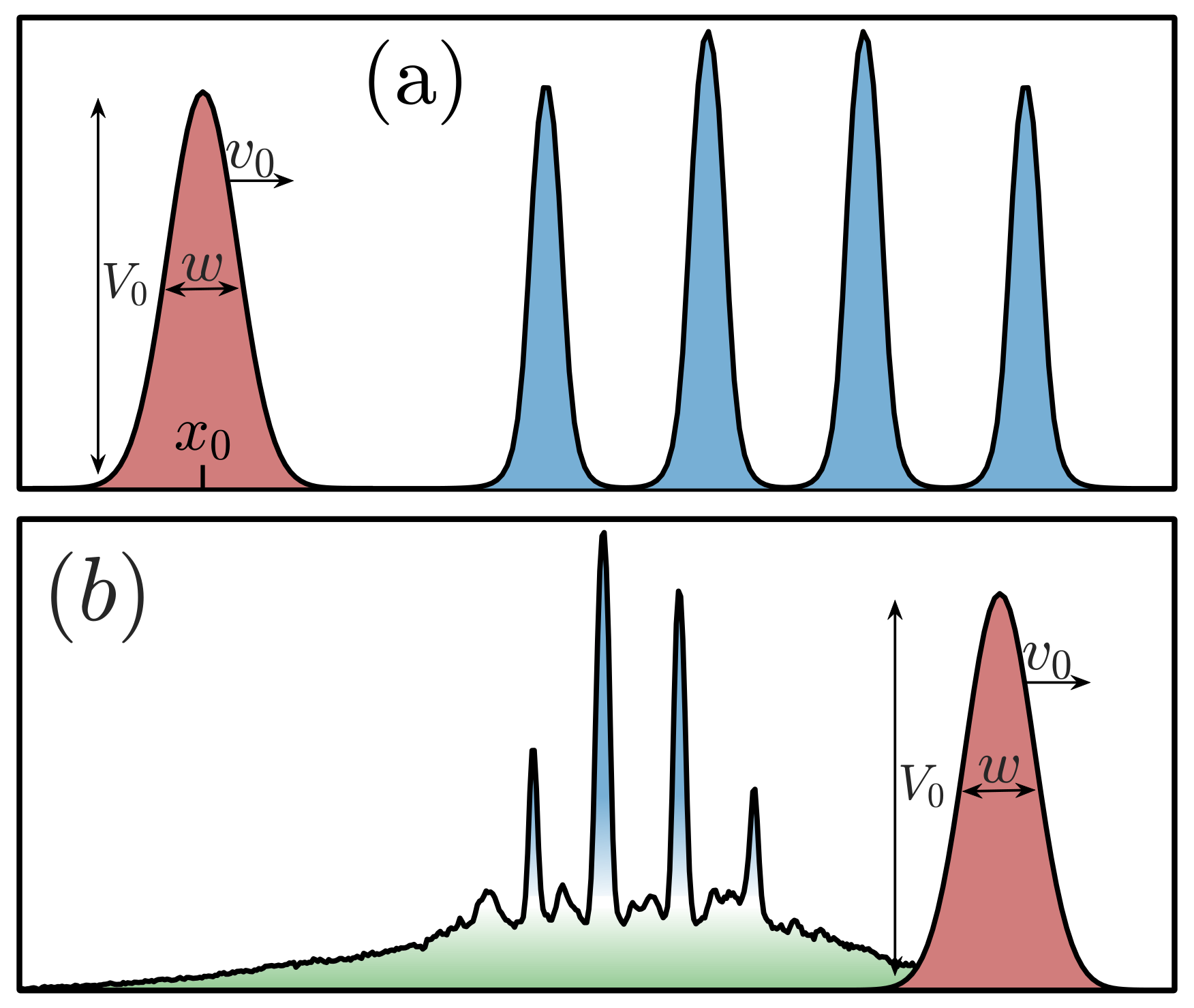}
\caption{(a) Schematic illustration of the driving protocol. A Gaussian barrier (red shaded region) located at $x_0$ and being characterized by height $V_0$ and waist $w$, is swept towards a quasi-1D array of $^{164}$Dy droplets (blue shaded area).
(b) The barrier sweep results in the dynamical generation of a supersolid, exhibiting an excited superfluid background, denoted by the green shading, on top of which crystalline structures persist.
}
\label{fig:schematic}
\end{figure}

We consider a quasi-1D array of dipolar droplets at zero temperature, see also Fig.~\ref{fig:schematic}(a). The relative interaction ratio between long- and short-range coupling is tuned to be larger than unity (see also below) such that we enter the droplet regime~\cite{chomaz_dipolar_2022}.  
The ensuing droplet array consists of $N=8\times 10^4$ $^{164}$Dy atoms, all of them polarized along the $z$ axis with the aid of a magnetic field. The array is confined by means of a harmonic potential with the following trap frequencies $(\omega_x,\omega_y,\omega_z)= 2\pi \times (19,53,81)~\rm{Hz}$, yielding an elongated cigar-shaped geometry similarly to the recent experiments of Refs.~\cite{bottcher2019transient,tanzi2019supersolid}.

\begin{figure*}[t!]
\centering\includegraphics[width=1.0\linewidth]{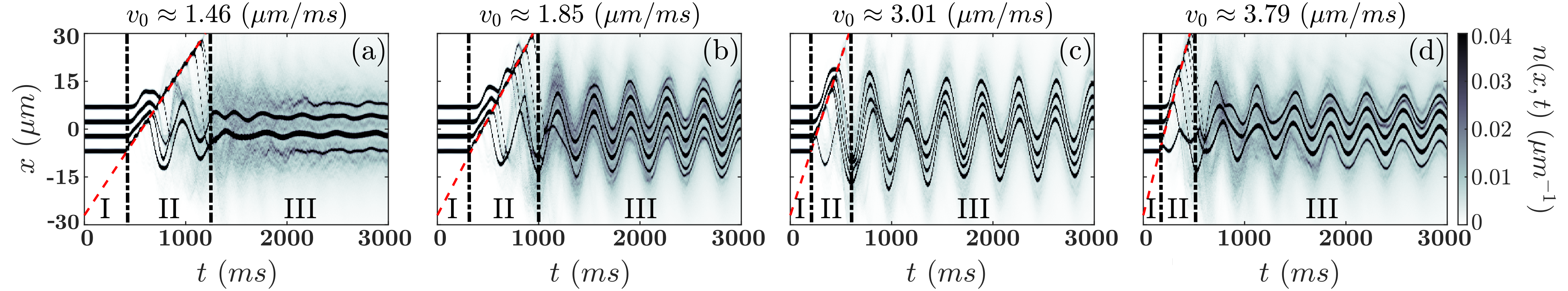}
\caption{Spatiotemporal 1D integrated density, $n(x,t)$, evolution of the droplet lattice consisting of four crystals after being  perturbed by a Gaussian barrier. The latter travels with different velocities (see legends) toward the droplet array. 
The barrier trajectory is illustrated by the red dashed line.
At long evolution times, following the crystal collisions, atom redistribution gives rise to a superfluid background atop which the post-collision   crystals oscillate in-sync, thereby manifesting the emergence of supersolidity. 
The vertical dash-dotted lines designate the three distinct dynamical regimes corresponding to the unperturbed crystal lattice (\rm{I}), the droplet collisions (\rm{II}), and the superfluid background nucleation (\rm{III}). 
In all cases, the characteristics of the Gaussian barrier correspond to  $V_0=10~\hbar \omega_x$, and $w=0.31~\mu m$, while the system consists of  $N=8\times 10^4$ $^{164}$Dy atoms with $a=86~a_0$ and $a_\text{dd} = 131~a_0$.}
\label{fig:Density_dynamics}
\end{figure*}

To determine the ground-state droplet configuration of the dBEC in the above quasi-1D geometry as well as track the emergent dynamics, an extended mean-field description is adopted~\cite{Santos2016filemanets,bisset_ground-state_2016,chomaz2016quantum,ferrier-barbut_observation_2016,baranov_condensed_2012}, see also Appendix~\ref{app:simulations}. 
This has been shown to be capable of adequately addressing different superfluid, supersolid and droplet phases of the dipolar gas~\cite{Ronen_dipolar_2006,Bortolotti_scattering_2006}.
In particular, the dipolar quantum gas is described by a macroscopic wave function $\Psi(\boldsymbol{r},t)$, obeying the following extended Gross-Pitaevskii equation (eGPE)~\cite{chomaz_dipolar_2022},
\begin{gather}
i \hbar \partial_t \Psi(\boldsymbol{r},t) = \Bigg [ - \frac{\hbar^2}{2m} \nabla^2 +V(\boldsymbol{r},t)  + \frac{4\pi \hbar^2 a}{m} \abs{\Psi(\boldsymbol{r},t)}^2  \nonumber \\
+ \int \text{d}^3\boldsymbol{r'}~ U_\text{dd}(\boldsymbol{r} -\boldsymbol{r'})\abs{\Psi(\boldsymbol{r'},t)}^2  \nonumber \\
+  \gamma (\epsilon_\text{dd}) \abs{\Psi(\boldsymbol{r},t)}^3 \Bigg ] \Psi(\boldsymbol{r},t), 
\label{Eq:MF_equation}
\end{gather}
where $\boldsymbol{r}=(x,y,z)$, $\hbar$ is the reduced Planck constant, and $m$ is the atomic mass. 
The first term refers to the kinetic energy, and the second to the external  confinement experienced by the atoms, see also the following discussion.
An important aspect of dipolar gases is the competition between both short-range (third term) and long-range (fourth term) interactions. The short-range interaction is characterized by the scattering length $a$, tunable by means of Fano-Feshbach resonances~\cite{Maier_broad_2015,Chin_Feshbach_2010}.
Here, in order to achieve the quasi-1D  droplet array, a scattering length $a=86~a_0$ is employed, with $a_0$ representing the Bohr radius.
On the other hand, the dipolar long-range anisotropic interaction potential reads 
\begin{equation}
U_\text{dd}(\boldsymbol{r}) = \frac{3\hbar^2 a_\text{dd}}{m}\left[\frac{1-3\cos^2\theta}{\boldsymbol{r}^3}\right]. 
\label{Eq:Dipolar_interactions}
\end{equation}
Apparently, it depends on the angle $\theta$ between the line connecting two dipoles and the $z$ axis.
In contrast to the short-range  scattering length, the dipolar length of $^{164}$Dy is fixed to $a_\text{dd} = 131~a_0$. 
In our setup, ground-state droplet configurations exist in the short-range interaction regime $a < 90~ a_0$, while supersolids [superfluids] take place for $a \in (90,100)~a_0$ [$a>100~a_0$]. 
The ground states of our system are obtained via imaginary
time propagation method of the eGPE [Eq.~(\ref{Eq:MF_equation})]; for further details of this approach see Appendix~\ref{app:simulations}. 

The final term in Eq.~\eqref{Eq:MF_equation} is instrumental in properly stabilizing high-density structures such as droplet arrays or supersolids, both of them occurring at large interaction strength ratios, $\epsilon_{\rm{dd}}= a_{\rm{dd}}/a$.
Such a contribution stems from the LHY correction to the dipolar mean-field energy functional within the local density approximation~\cite{Lima_quantum_2011,schutzhold_mean_2006}.
The relevant coefficient is explicitly given by   $\gamma(\epsilon_\text{dd})=\frac{128 \hbar^2\sqrt{\pi}a^{5/2}}{3m}  \left(1+\frac{3}{2} \epsilon^2_\text{dd}\right)$~\cite{Lima_quantum_2011}.

To probe the dynamical response of the droplet crystal and nucleate supersolidity, the following driving  protocol is devised.
A Gaussian barrier, characterized by height $V_0$ and width $w$, is dragged towards the droplet array with speed $v_0$ [see also the schematic in Fig.~\ref{fig:schematic}(a)].
Initially, the barrier is placed far away from the array, at position $x_0=-27 ~ \mu m$, so that no overlap occurs between the beam and the atoms.
The moving barrier can be experimentally realized by shining and subsequently sweeping a focused laser beam~\cite{meinert2017bloch,Engels_barrier}, hence modifying the trapping potential of the dipolar atoms.
Explicitly, the time-dependent external potential takes the form 
\begin{equation}
V(\boldsymbol{r},t) =  \sum_{\substack{k=x,\\y,z}} ( \frac{m \omega_k^2 k^2 }{2} ) 
+ ~V_0~\exp\Bigg\{   - \frac{\left[ x - (x_0+v_0t)  \right]^2  }{2w^2} \Bigg\}.
\label{Eq:Gaussian_barrier}
\end{equation}
Since our ultimate goal is to significantly perturb the droplet lattice, the barrier height and width are chosen to be comparable to the chemical potential, $\abs{\mu} \simeq 30 ~ \hbar \omega_x$, of the ground-state configuration and the full-width-at-half-maximum, $\sigma$, of the droplet peaks respectively, $\sigma = 0.9 ~ \mu m$. 
For this reason, we use  $V_0=10~\hbar\omega_x$, and $w=0.31~\mu m$. Moreover, the selected barrier speeds are comparable 
to the velocity scale set by the chemical potential of the droplet lattice in its ground-state, namely $v_g=\sqrt{\abs{\mu}/m} = 1.17 ~ \mu m/ms$.

\begin{figure}[ht]
\centering\includegraphics[width=1\linewidth]{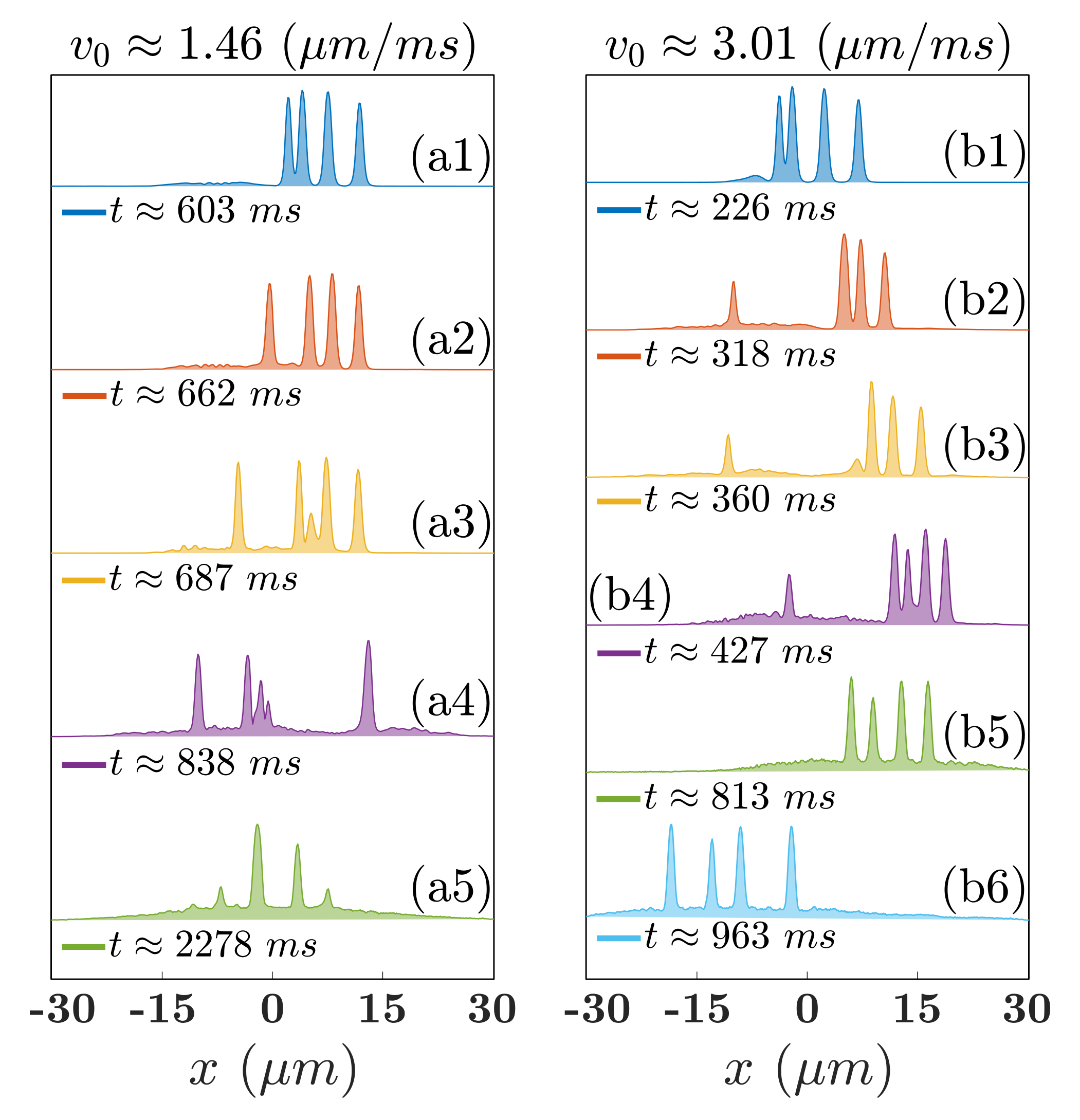}
\caption{Droplet inelastic collisions induced by the barrier sweep at selected time instants of the density evolution (see legends) yielding the appearance of a superfluid background. The integrated 1D density profiles, $n(x,t)$, have been rescaled and shifted appropriately to ensure a proper visualization. The barrier velocity refers to (a1)-(a5) $v_0  \approx 1.46 ~ \mu m/ ms$, and (b1)-(b6) $v_0 \approx 3.01 ~ \mu m/ms$, see also Fig.~\ref{fig:Density_dynamics}(a), (c). All other parameters are the same as in Fig.~\ref{fig:Density_dynamics}.}
\label{fig:Density_cuts}
\end{figure}

\section{Barrier sweep induced supersolidity}\label{sec:dynamics}

The above-discussed dynamical protocol leads to a response quite distinct from just exciting the normal modes of droplets, a process that can be modeled by coupled oscillators as reported in Refs.~\cite{Mukherjee_classical_2023,Bougas_signatures_2025}.
Specifically, it will be demonstrated that when the Gaussian barrier approaches the droplet crystal it perturbs the latter giving rise to droplet collisions accompanied by  significant  atom and energy redistribution. 
For longer evolution times, where the barrier is outside of the crystal lattice, the aforementioned processes result in an appreciable superfluid background beneath the droplets. 
This clearly hints toward the creation of a supersolid configuration [Fig.~\ref{fig:schematic}(b)] notably in the interaction regime where ground-state droplet states exist.

Such a mechanism is easily visualized by inspecting the integrated 1D densities, $n(x,t)=\int dydz~ \abs{\Psi(\boldsymbol{r},t)}^2$, that can be experimentally traced through single-shot in-situ measurements~\cite{Sohmen_SS_2021}. 
The corresponding density evolution for different velocities of the barrier are presented in Fig.~\ref{fig:Density_dynamics}. 
Apparently, three distinct dynamical stages can be discerned, denoted by the roman letters in Fig.~\ref{fig:Density_dynamics}.  
The first one (\rm{I}) refers to early evolution times, where the Gaussian barrier (its trajectory denoted by the tilted red dashed line in Fig.~\ref{fig:Density_dynamics}) has not yet reached the first ($x \simeq -7 ~ \mu m$) crystal of the droplet array. 
Such a regime occurs within the interval $t \leq t^* = \abs{\Delta x}/v_0$, where $\abs{\Delta x} = 19.5 ~ \mu m$ is the distance between the initial barrier's position and the nearest droplet peak.  
Naturally, within this time interval the droplet array remains stationary irrespective of the barrier velocity. 
We remark that the initial droplets do not remain stationary in the presence of three-body recombination. Instead, they drift toward the trap center and lose population already within this first dynamical stage, as discussed in Appendix~\ref{app:3body}.

\subsection{Droplet collisions}  \label{sec:collisions}

The second stage (\rm{II}), which is drastically more complex compared to the first one,  involves the head-on collision of the Gaussian barrier with the crystals, leading to subsequent interactions between the individual droplets, see also the density snapshots shown in  Fig.~\ref{fig:Density_cuts}.
The barrier first collides with the nearest ($x \simeq -7 ~ \mu m$) crystal of the droplet array, which becomes excited, emitting particles in the form of a small density trail attached to the left of the first droplet [Fig.~\ref{fig:Density_cuts} (a1), (b1)].
This process alludes to the self-evaporation mechanism present in dipolar quantum droplets~\cite{Baillie_collective_2017}.

Subsequently, the excited first droplet forms a collisional complex with the second droplet, characterized by a significant in-between density overlap, facilitating particle tunneling~\cite{Biagioni_measurement_2024,donelli_self_2025} among the crystals.
Eventually, the first droplet recoils to the left [Fig.~\ref{fig:Density_cuts}(a2), (b2)] due to the repulsion experienced by dipolar droplets colliding along the direction transverse to their polarization axis~\cite{Adhikari_statics_2017,Ferrier_liquid_2016}.
After the collision the amplitude of the participating droplet peaks changes, thereby highlighting 
the aforementioned tunneling processes [Fig.~\ref{fig:Density_cuts}(b2)].

The second droplet is now excited both due to the ensuing inelastic collisions with the first droplet, but also because of the emitted particles originating from self-evaporation of the first droplet. Moreover, the kick provided by the latter is sufficient to lead to consecutive collisions of the second droplet with the rest of the crystals in the  droplet array.
An avalanche mechanism thus takes place, where the droplet crystals become further excited by the emitted particles from the other droplets (owing to their  self-evaporation) and also by the tunneling induced by the inelastic collisions between the crystals  [Fig.~\ref{fig:Density_cuts} (a3), (a4), (b3) and (b4)].
The imprint of the above-discussed  tunneling processes is the change in the amplitude of the involved droplets during their interactions.

\begin{figure}[t!]
\centering\includegraphics[width=0.9\linewidth]{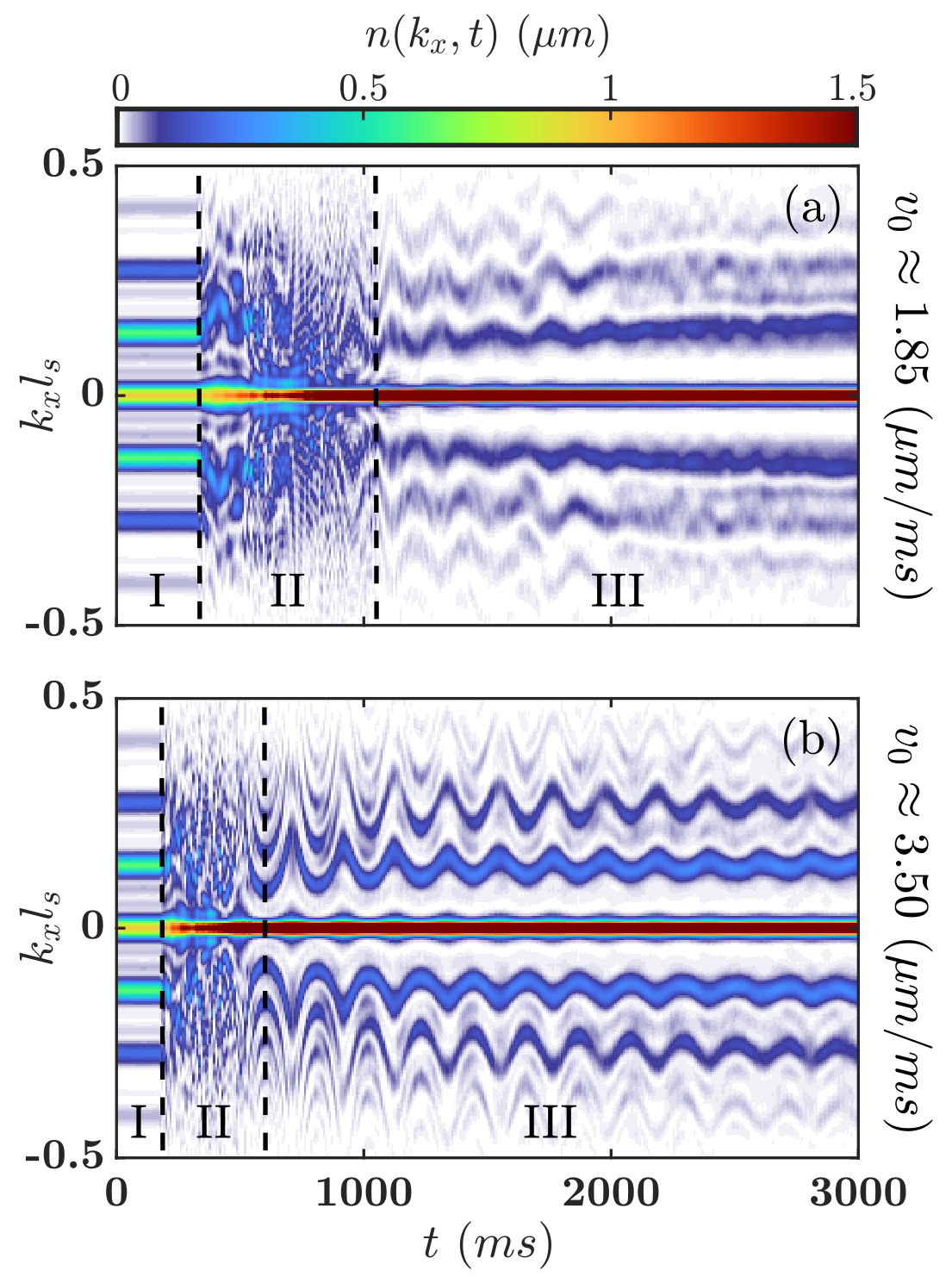}
\caption{Time-evolution of the 1D momentum distribution for two different barrier velocities (see legends). 
The gradual formation of a superfluid background at long evolution times becomes evident from the underlying momentum transfer toward the zero momentum peak. 
This process is associated with the dynamical generation of the supersolid. 
The vertical dashed lines delineate the same dynamical regimes, i.e. \rm{I}, \rm{II} and \rm{III}, identified in Fig.~\ref{fig:Density_dynamics}.
The remaining system parameters are the same as in Fig.~\ref{fig:Density_dynamics}.}
\label{fig:momentum}
\end{figure}

In the third stage (\rm{III}) occurring at long evolution times, e.g. $t>1500 ~ ms$, the agglomeration of all the emitted particles stemming from the inelastic droplet collisions results in the development of a significant superfluid background [Fig.~\ref{fig:Density_cuts}(a5), (b5) and (b6)]. This superfluid background persists for longer evolution times, even in the presence of experimentally relevant three-body losses, see also Appendix~\ref{app:3body}.
At the same time, the individual droplet peaks oscillate in-sync, almost free of any inelastic collisions [Fig.~\ref{fig:Density_dynamics}]. This  hints towards the establishment of phase coherence among them, which is a central property of supersolids~\cite{Ilzhofer2021}. 
Both the in-sync droplet motion and the superfluid fraction are affected by the barrier velocity.
The impact of the latter refers mostly to the degree of excitation of the superfluid substrate [c.f. Fig.~\ref{fig:Density_dynamics}(a) and (c)] and the amplitude of the collective droplet excitations [c.f. Fig.~\ref{fig:Density_dynamics}(c) and (d)].
The effect of $v_0$ will become more evident in the following, see in particular  Fig.~\ref{fig:SF_fraction}(a) and also the discussion in  Sec.~\ref{sec:SF_fraction}.

It is important to note that the formation of the superfluid background occurs for large barrier heights, $V_0 \gtrsim 7 ~ \hbar \omega_x \sim \abs{\mu}/4$, see also the discussion in Sec.~\ref{sec:SF_fraction} and Fig.~\ref{fig:SF_fraction}(b). 
For smaller $V_0$ the interaction of the barrier with the droplets is weaker, and as a consequence the droplet array undergoes a collective motion similar to the normal modes of a 1D crystal chain~\cite{Mukherjee_classical_2023,Bougas_signatures_2025}.
In contrast, for $V_0 \gtrsim \abs{\mu}$ the kinetic energy imparted by the barrier to the array is comparable and larger than the binding energy of the array, and hence the barrier drag entails melting of the droplet lattice.

\subsection{Superfluid background nucleation} \label{sec:superfluid}

To further support the emergence of a supersolid configuration caused by the barrier sweep, we next invoke additional observables in order to justify the creation of the superfluid background and the rigidity of the crystals. 
A relevant measure is the 1D momentum distribution, $n(k_x,t) = \abs{\int dx ~ e^{-i k_x x} \sqrt{n(x,t)} }^2$ that can be experimentally  obtained via time-of-flight imaging~\cite{Chomaz2018}. 
The dynamics of $n(k_x,t)$ is depicted in Fig.~\ref{fig:momentum} for two paradigmatic barrier velocities.
It can be readily seen that initially ($t=0$), a well-defined series of momentum peaks is significantly populated in a hierarchical order as a result of the existent lattice structure of the droplet array, see also Fig.~\ref{fig:schematic}(a).
In particular, the first prominent sideband occurring at $k_x l_s \simeq \pm 0.13$ in Fig.~\ref{fig:momentum}(a) and (b) refers to the lattice spacing, stemming from the roton minimum present in the excitation spectrum of the dipolar gas~\cite{petter_probing_2019}. 
Note that multiples of that peak appear with smaller amplitude, which is  attributed to the deviations of the droplet array from a regular lattice. 
Here, $l_s = \hbar/ \sqrt{m \abs{\mu}}$ is the characteristic length scale set by the chemical potential of the initial state~\cite{Bougas_generation_2026}.

Upon the interaction of the barrier with the droplet array (onset of stage {\rm II}) the momentum peaks become distorted, a behavior that reflects the loss of the lattice structure due to the inter-droplet collisions.
Note that this dynamical response interval depends on the velocity of the barrier, since the latter collides with the droplets at time instants which are inversely proportional to $v_0$. As such, it occurs within the interval $t \in (340,1051)~ms$ [$t \in (186,601)~ ms$] for $v_0=1.85~ \mu m/ms$ [$v_0=3.5 ~\mu m/ms$] as shown in Fig.~\ref{fig:momentum}(a) [Fig.~\ref{fig:momentum}(b)]. 
At longer evolution times, e.g. $t>1000 ~ ms$ in Fig.~\ref{fig:momentum}(a), a regular series of momentum peaks appears again, displaying an oscillatory behavior.
This is attributed to the in-sync collective motion of the droplet peaks, see also Fig.~\ref{fig:Density_dynamics}, leading to a rearrangement of the crystal lattice structure, which in turn affects the lattice spacing.
Also, the almost constant peak amplitude of the individual momentum modes suggests the suppression of tunneling between the crystals mediated by the superfluid background within this dynamical stage {\rm III}.
Simultaneously, the amplitude of the zero momentum peak acquires substantial population, becoming the dominant momentum mode, a feature that persists at even longer evolution times.
Such an enhancement is linked to momentum transfer from higher- to lower-lying modes and importantly to the zero momentum mode. This further corroborates the development of a pronounced superfluid substrate, since the latter is always associated with a sharp zero momentum peak in Fourier space~\cite{Pitaevskii_Bose_2016}.

\begin{figure}[t!]
\centering\includegraphics[width=0.95\linewidth]{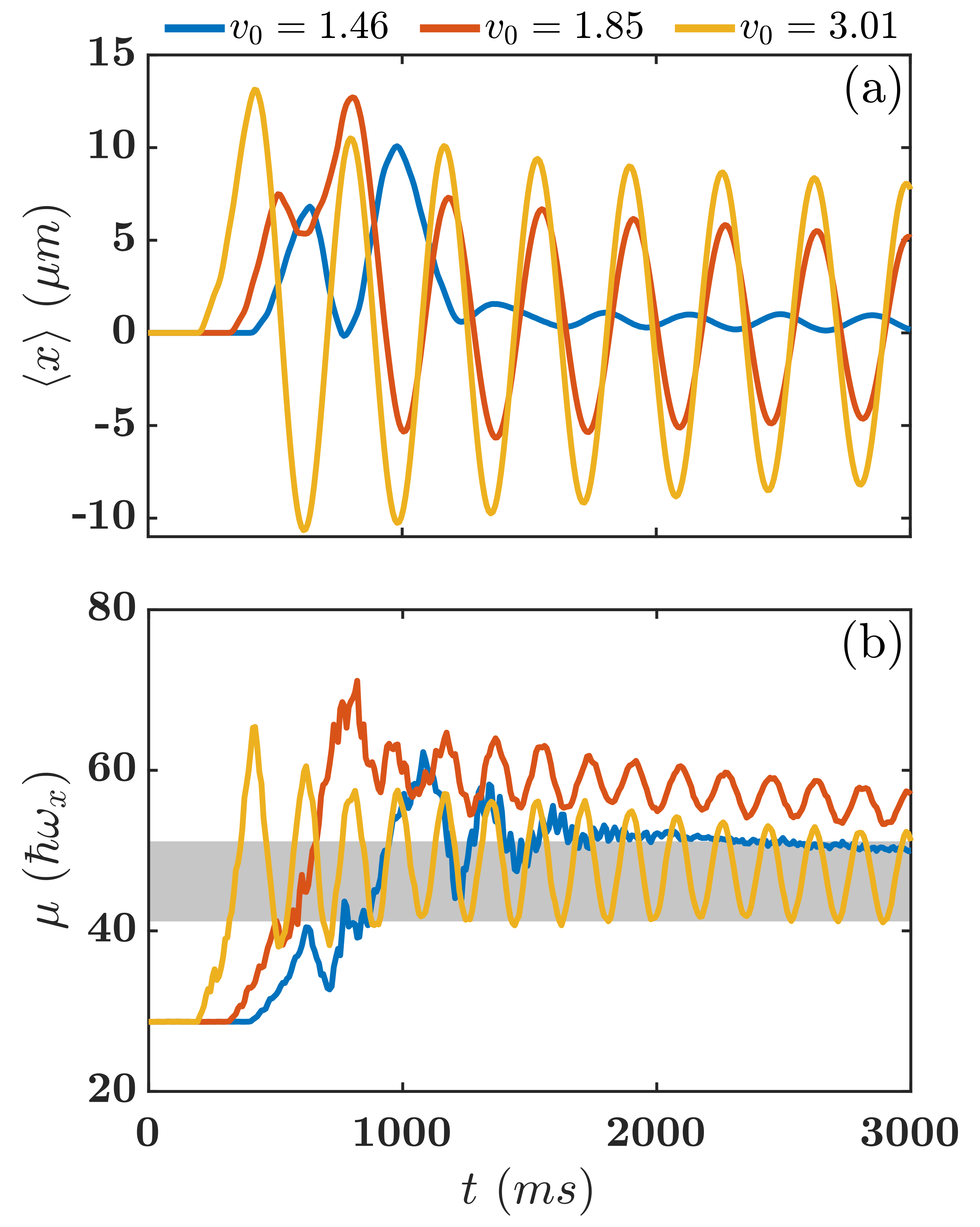}
\caption{(a) Dynamics of the center-of-mass of the dipolar gas, $\braket{x}$, for different barrier velocities (see legend) capturing the  collective motion triggered by the barrier sweep. This collective motion, in particular, occurs at evolution times where the barrier is long past the dipolar gas (regime \rm{III}). (b) The time-evolution of the associated chemical potential exhibits an increase (regime \rm{II}) and a subsequent oscillatory behavior (regime \rm{III}) unveiling underlying energy redistribution induced by the barrier drag. Afterwards, the chemical potential shows a saturation tendency toward a value larger than the respective $\mu$ of the initial droplet ground-state but in the ballpark of the supersolid ground-state phase marked by the shaded area. 
This occurs in the dynamical stage \rm{III}, where the superfluid background has been established and remains persistent.   
The parameters for the potential barrier and the dipolar gas are the same as in Fig.~\ref{fig:Density_dynamics}.}
\label{fig:Chemical_potential_disp}
\end{figure}

\subsection{Energy injection and center-of-mass motion} \label{sec:Energy}

Apart from the pronounced superfluid background another important dynamical feature pointing towards the nucleation of a supersolid is the in-sync motion of the droplets at later evolution times [Fig.~\ref{fig:Density_dynamics}], and in particular within stage {\rm III}.
In fact, such a motion resembles the dipole mode of the center-of-mass of dBECs and normal BECs, when displaced from the center of a harmonic trap~\cite{Dalfovo_theory_1999}.
To further quantify this collective behavior, we inspect the time evolution of the center-of-mass, $\braket{x(t)} = \int dx ~ x ~n(x,t)$, presented in Fig.~\ref{fig:Chemical_potential_disp}(a) for different velocities of the Gaussian barrier.
At long evolution times the frequency of the droplet peak oscillations is roughly $3~ \rm{Hz}$, regardless of the barrier velocity. 
The latter affects mainly the amplitude and phase of the oscillations.
In particular, for large $v_0$ the collective motion of the crystals  becomes more pronounced as inferred by the increased  amplitude of $\braket{x(t)}$ [Fig.~\ref{fig:Chemical_potential_disp}(a)].

The frequency of oscillations is much smaller than the trap frequency along the weakly confined direction, namely $\omega_x/(2\pi) = 19 ~ \rm{Hz}$.
Therefore, the aforementioned in-sync oscillations can not be solely attributed to the dipole motion of the center-of-mass, whose frequency should coincide with $\omega_x/(2\pi)$ in this quasi-1D geometry.
Instead, the dynamics of $\braket{x(t)}$ alludes to the presence of other low energy modes~\cite{Mukherjee_classical_2023}.
Low-frequency collective modes are actually absent in normal superfluids, but rather exist in dipolar supersolids~\cite{Guo2019,Hertkorn_decoupled_2024}, e.g. the out-of-phase Goldstone mode.
Such low-frequency collective modes can lead to a decrease in the time period of the center-of-mass oscillations, as reported in the dynamical response of supersolids driven out-of-equilibrium by means of momentum kicks~\cite{Mukherjee_classical_2023}.
In addition, the decay of the center-of-mass oscillations observed in Fig.~\ref{fig:Chemical_potential_disp}(a) can also be attributed to the presence of low-energy collective modes~\cite{Mukherjee_classical_2023}, and to the energy exchange between the crystal and the superfluid background.
Note that, in the case of a potential defect dragged through a dipolar superfluid, there is no decay of the center-of-mass oscillations, and the oscillation frequency is approximately $\omega_x/(2\pi)$ [not shown for brevity]. 
This further supports our argument for the existence of low energy modes in our dynamically created supersolid.

The decay of the center-of-mass oscillations in the course of the evolution, underscores the importance of energy redistribution  processes in the dynamical generation of a supersolid.
To further understand such processes, we resort to the evolution of the chemical potential $\mu(t)$ for different velocities [Fig.~\ref{fig:Chemical_potential_disp}(b)].
The chemical potential roughly refers to the total energy per particle and provides further information regarding the energy injection from the barrier defect to the droplet array, as well as the in-sync oscillations of the latter at later evolution times. 
At initial times within the first (pre-collision) dynamical stage \rm{I}, $\mu(t)$ is constant, reflecting the invariance of the droplet array prior to the interaction with the barrier, see also Fig.~\ref{fig:Density_dynamics}.
Of course, the duration of such a regime depends inversely proportional to $v_0$, since it occurs up to $t \leq t^*$.

At the onset of stage \rm{II}, after the Gaussian defect hits the first droplet peak, a steep linear increase occurs in the chemical potential. 
The slope of such a linear response depends on $v_0$, with larger slopes associated with faster moving barriers, which impart a larger amount of kinetic energy to the droplet array.
During the stage related to the droplet collisions (see region \rm{II} in Figs.~\ref{fig:Density_dynamics} and~\ref{fig:momentum}), a nonlinear response is observed in $\mu(t)$, which then transitions to an oscillatory behavior at late evolution times.
Such a response refers to the third dynamical stage (${\rm III}$) related with the generation of a supersolid. The oscillations feature a decaying trend, and the damping mechanism is the same as the one responsible for the damping of the center-of-mass oscillations [Fig.~\ref{fig:Chemical_potential_disp}(a)], i.e. the energy exchange between the droplet peaks and the underlying superfluid background. 
The mean values of $\mu(t)$ feature a saturation tendency toward the chemical potential region associated with supersolids in the ground-state of our system. 
Such a region is marked by the shaded area in Fig.~\ref{fig:Chemical_potential_disp}(b). 
Importantly, this is achieved dynamically due to the driving protocol while keeping the interactions fixed corresponding to the ground-state droplet region.

Notice that $\mu(t> 2000 ~ ms)$ is slightly larger than the respective chemical potential values in the supersolid phase of the ground-state.
This behavior is traced back to the fact that the dipolar gas is excited due to its collision with  the potential barrier. Even if it acquires supersolid characteristics, its energy and chemical potential are (at least slightly) larger than the ones corresponding to the respective supersolid ground-state phases. 
Regarding the individual energy contributions, the LHY and dipolar energies both decrease 
in magnitude at late evolution times, when the system enters stage {\rm III}  (not shown for brevity).
Such a trend provides a further hint towards the establishment of a supersolid state, which is characterized by a smaller contribution of the dipolar and LHY energies as compared to the isolated droplets.
Moreover, the barrier collision with the droplet array results in the enhancement of the kinetic energy of the dipolar gas, since the center-of-mass of the latter is set to a persistent motion.

\subsection{Superfluid fraction}  \label{sec:SF_fraction}

To understand the dynamical generation of a supersolid in a more systematic way, we employ the superfluid fraction as a relevant order parameter.
In particular, we utilize the upper bound of the superfluid fraction~\cite{Kirkby_kibble_2025,leggett_can_1970,Biagioni_measurement_2024} which reads 
\begin{equation}
    f_s^u = 2L^2  \left(  \int_{-L}^L \frac{dx}{n(x,t)}  \right)^{-1}.
    \label{Eq:Superfluid_fraction}
\end{equation}
Note that the above measure pertains to gases set in rotation in annular geometries\footnote{or atom chains with periodic boundary conditions and a phase twist~\cite{Biagioni_measurement_2024,Matos_transitional_2025}.}~\cite{leggett_can_1970}, and therefore it is employed here only in a heuristic sense.
The main advantage of such a measure are the well-defined bounds of the homogeneous  superfluid and isolated droplet  phases, where $f_s^u=1$ and $f_s^u=0$ respectively.
Notably, for supersolids it holds that $0<f_s^u<1$, depending on the amplitude of the underlying superfluid background. 

Since our dipolar gas is confined in a harmonic trap, which enforces a density curvature even in the superfluid phase, the upper bound of $f_s^u$ is different than unity~\cite{chauveau_superfluid_2023}. 
Moreover, it depends on the choice of the $[-L,L]$ spatial integration  interval.
Here, we choose $L=20~\mu m$ which is able to cover the entire spatial extent of the ground-state configuration (including all crystals) and can accommodate at least two droplet peaks during the evolution, see also Fig.~\ref{fig:Density_dynamics}. 
For such a choice, the superfluid fraction of three representative ground states, of our harmonically trapped dBEC, corresponding to fully formed droplets, supersolids and superfluids at $a=86,~96,~ 130~a_0$ is $f_s^u=0,0.2,0.4$ respectively.
Hence, it is evident that within this length scale set by $L$, droplets are always characterized by $f_s^u=0$, while supersolids display a superfluid fraction larger than zero. 
In fact, $f_s^u = 0.4$ in the ground-state transition region from supersolid to superfluid of our setup occurring at $a = 100 ~ a_0$.
Hence, $0 < f_s^u< 0.4$ indicates the dynamical nucleation of supersolids in our setting~\cite{bombin_quantum_2023}.

\begin{figure}[t!]
\centering\includegraphics[width=1\linewidth]{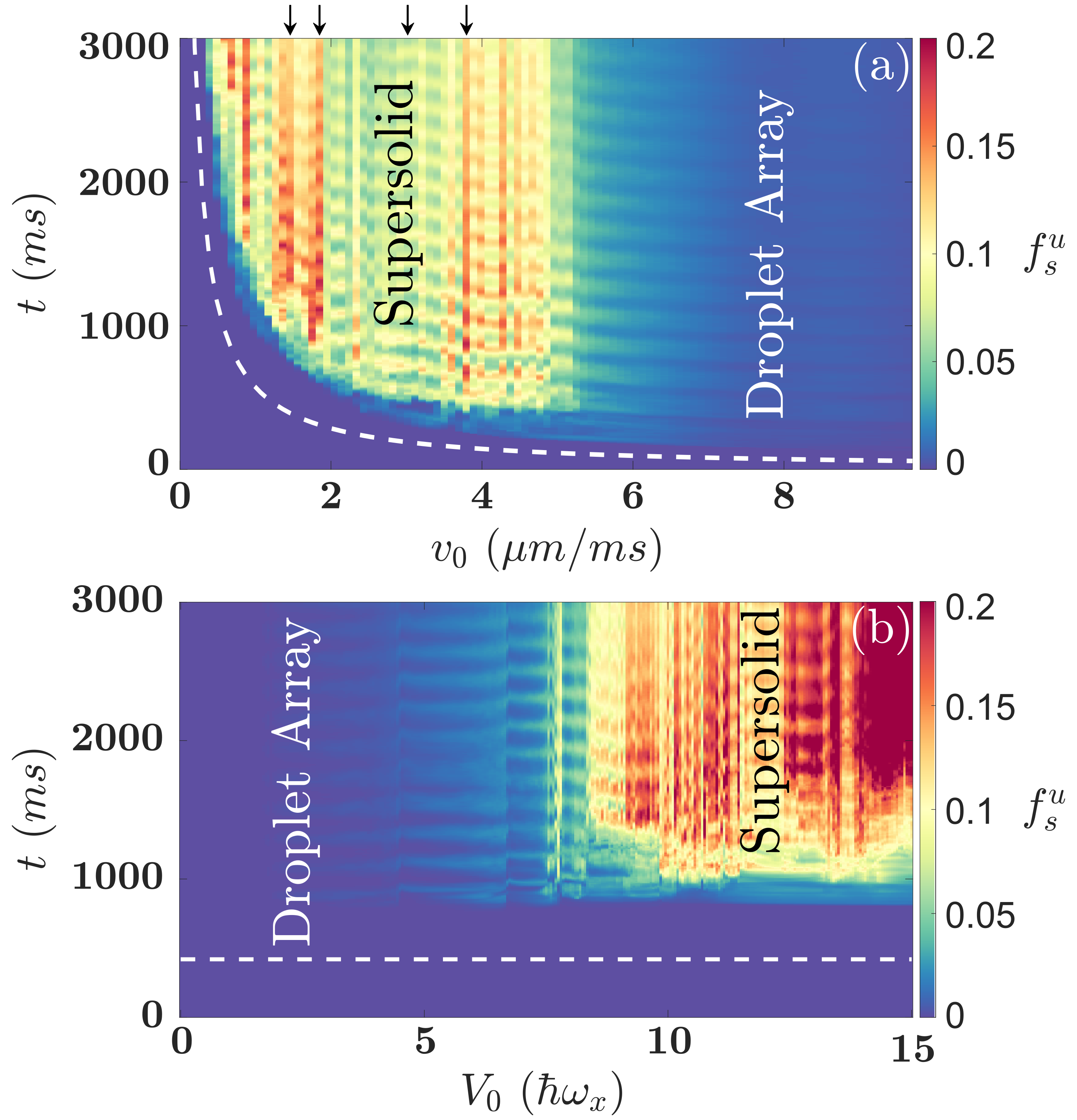}
\caption{Time-evolution of the superfluid fraction, $f_s^u$ [Eq.~(\ref{Eq:Superfluid_fraction})] with respect to (a) the barrier velocity and $V_0=10 ~ \hbar \omega_x$ fixed as well as (b) the barrier height and constant velocity $v_0=1.36~ \mu m/ ms$. 
The ${}^{164}$Dy dipolar atoms, $N=8\times 10^4$, are initialized in a droplet array with four crystals being subjected to a Gaussian barrier sweep with velocity $v_0$. This process leads to the nucleation of a supersolid, as inferred by the increase of the superfluid fraction in the course of the evolution within the velocity range $v_0 \in (0,6)~ \mu m / ms$ in panel (a) and the height interval $V_0 \in (7,15)~ \hbar \omega_x$ depicted in panel (b). These parametric regions are marked as ``Supersolids" in panels (a), (b). The white dashed line in both panels corresponds to $t=t^*$, delineating the first instant that the barrier collides with the outermost crystal. The vertical arrows in panel (a) mark selected barrier velocities, displayed in Fig.~\ref{fig:Density_dynamics}. Other system parameters are the same with the ones of Fig.~\ref{fig:Density_dynamics}.}
\label{fig:SF_fraction}
\end{figure}

The superfluid fraction therefore represents an order parameter during the dynamics resulting in the phase diagrams shown in Fig.~\ref{fig:SF_fraction}.  These underline the effect of the barrier velocity or height on the dynamical formation of supersolidity.
Focusing on the impact of the velocity for fixed height, see Fig.~\ref{fig:SF_fraction}(a), reveals that before the barrier hits the initial droplet array, the superfluid fraction is consistently zero regardless of the barrier velocity. 
This is the dynamical regime ${\rm I}$ occurring at $t \leq t^*$, see the white dashed line in Fig.~\ref{fig:SF_fraction}(a) representing $t=t^*$.

For evolution times $t>t^*$, two superfluid fraction response regimes can be distinguished depending on the barrier's velocity. 
More concretely, for $v_0 > 6 ~ \mu m / ms$ the fraction $f_s^u$ remains zero throughout the evolution including the dynamical stages ${\rm II}$ and ${\rm III}$.
This is the diabatic driving  regime, where the barrier moves much faster than the characteristic velocity set by the chemical potential of the droplet array in its ground-state, i.e. $v_0 \gg v_g$.
As a result, the dipolar gas remains almost entirely unperturbed even at significantly longer evolution times following the collision with the barrier.

In contrast, for $v_0 \lesssim v_g$ the superfluid fraction becomes finite and oscillatory for $t>t^*$, characterized by small amplitude and multi-frequency oscillations.
In fact, the time-averaged 
value of  $f_s^u$ within stage {\rm III} becomes comparable to the superfluid fraction pertaining to the stationary supersolid states of our setup taking place at scattering lengths $a \in [92,97] ~ a_0$ and  characterized by $f_s^u = [0.08,0.3]$.
Such a behavior is further supported by the long-time evolution of the chemical potential [Fig.~\ref{fig:Chemical_potential_disp}(b)].
Recall that the long-time damped
oscillations of $\mu(t)$ take place around the chemical potential values referring to the supersolid ground states, see the shaded area in Fig.~\ref{fig:Chemical_potential_disp}(b). Overall, the superfluid fraction becomes finite after the barrier is dragged through the droplet array, namely at the end of dynamical stage ${\rm II}$ and through the entire regime ${\rm III}$, retaining its finite value even at late evolution times.
Such a behavior, in conjunction with all the other measures presented above, is strongly suggestive of the dynamical generation of supersolids out of isolated droplets. 
We once more emphasize that this phenomenon takes place within the interaction regime
where the dipolar gas supports only isolated droplets as its ground states.

Note that at some specific barrier velocities the superfluid fraction gets more enhanced compared to other $v_0$, pointing towards the existence of resonant features.
The latter appear also in the case of a droplet lattice with a different number of droplet peaks, see Appendix~\ref{app:different_crystals}. 
Interestingly, the evolution of the chemical potential with respect to the barrier velocity, $\mu(t)$, exhibits a similar dynamical behavior at the different velocity regimes. 
This implies that the aforementioned velocity dependent dynamical regions are related to energy redistribution processes of the dipolar gas after the barrier sweep, where in particular the LHY and dipolar interaction contributions decrease in magnitude followed by an increase of the kinetic energy. 
Another connection appears between the resonance patterns of $f_s^u$ and the Fourier spectra of the center-of-mass oscillations.
In the case of resonant enhancement of the superfluid fraction, the Fourier distributions are usually broader with many peaks.
Such an observation suggests the selective excitation of collective modes of the dipolar gas when the barrier collides with the latter at specific velocities. 
However, such an explanation is certainly not conclusive and further investigation, e.g. in terms of the Bogoliubov excitation spectrum within the supersolid ground-state regime, is required in future studies. 
This could also possibly allow to match specific excitation branches with the frequencies participating in the oscillatory behavior of $f_s^u$.

For completeness, we remark that the same resonant features and overall response of $f_s^u$ appear also when employing another order parameter for identifying supersolid formation, namely the density contrast~\cite{ripley_two-dimensional_2023,poli_excitations_2024}.
The latter yields the contrast between the high and low density segment of a modulated phase, being unity in the case of isolated droplets, and acquiring values smaller than unity for supersolids.
In that sense, the density contrast is complementary to the superfluid fraction providing an independent verification for the build-up of supersolidity and the overall response observed in  $f_s^u$.

Next, we focus on the impact of the barrier height at fixed $v_0$, see Fig.~\ref{fig:SF_fraction}(b).
The latter sets the $t^*$ time, before which the droplet array remains unperturbed (stage {\rm I}), as evidenced by the zero superfluid background below the white dashed line in Fig.~\ref{fig:SF_fraction}(b).
For $t>t^*$ two dynamical regimes of $f_s^u$ in terms of $V_0$ can be discerned, similar to the response of $f_s^u$ with respect to the barrier velocity.
First, for $V_0 \lesssim 7 ~ \hbar \omega_x$ the barrier imparts only a small amount of energy to the ${}^{164}$Dy atoms, since the barrier height is only a small fraction of the droplet's chemical potential, $V_0 \lesssim \abs{\mu}/4$.
As a result the initial state is weakly perturbed, and the dynamical response refers mostly to collective droplet modes, akin to those appearing in coupled oscillators~\cite{Bougas_signatures_2025,Mukherjee_classical_2023}.

In contrast, for $V_0 \gtrsim 7 ~ \hbar \omega_x$ the droplets become substantially  excited, leading to the generation of an excited superfluid background.
As a result, the superfluid fraction increases significantly over time, designating the dynamical nucleation of a supersolid.
Similar to the response with respect to the barrier velocity, $f_s^u$ becomes resonantly enhanced at particular $V_0$. 
However, the resonance patterns are quite distinct in both of these cases, implying that a moving defect is a unique probe for unveiling the excitation spectrum of dipolar gases~\cite{Platt_supersolid_2025}.
Again, such resonance features hint towards energy redistribution processes and selective excitation of collective modes, requiring a more systematic study in the future.
Note that when $V_0 > 15~ \hbar \omega_x$, the barrier height starts to become comparable to the binding energy of the droplet array, $\abs{\mu}$, and therefore the initial state is strongly perturbed.
The crystal can even melt completely, and any signatures of supersolidity are lost (not shown).

\section{Summary and Perspectives}\label{sec:conclusions}

We have investigated the nonequilibrium quantum dynamics of an elongated harmonically trapped droplet lattice of dipolar Dysprosium atoms subjected to a repulsive moving potential. 
To simulate the ground-state and the dynamical response of this setting, we deploy the suitable extended 
Gross-Pitaevskii model that encompasses first-order quantum corrections to the mean-field energy functional. 
The initial ground-state of the droplet lattice is attained by tuning the relative strength ratio between dipolar
and short-range interactions in the regime where the former dominate. 

The characteristics of the potential barrier, such as the width and the height, are chosen such that a significant amount of energy is injected to the initial droplet array. 
Consecutively, it is let to sweep with an initial velocity through the incoherent droplet lattice triggering collision events among the ensuing crystals. 
This process favors significant particle redistribution  eventually leading to the build-up of an excited superfluid background. 
The time-evolution of the chemical potential highlights the crucial role of energy injection and redistribution between the crystalline and superfluid components in the course of the dynamics.

The dynamical emergence of supersolidity is corroborated by the following complementary processes. 
First, monitoring the underlying momentum distribution during the evolution we identify a transfer from higher momentum modes to the zero-momentum peak. 
Second, the post-collision  crystals after the barrier sweep undergo a collective oscillatory motion while maintaining their shape for long evolution times, which is suggestive of their rigid character. 
Here, the oscillation frequency of the crystals is much smaller than the trap frequency along the elongated direction, which is indicative of the presence of low energy modes characteristic of supersolids. 
Moreover, we evaluate the associated superfluid fraction and observe its gradual development. 
This observable reveals that supersolid formation is dictated by the velocity and height of the potential barrier, leading to distinct dynamical response regimes.  
At moderate velocities with fixed height, the superfluid fraction is maximal featuring a non-monotonic 
behavior with the barrier velocity, while for increasing velocities the droplet lattice remains largely un-perturbed and coherence is mainly suppressed. 
Otherwise, for fixed velocity and varying height the superfluid fraction becomes prominent for heights comparable to the chemical potential of the initial state.

Our results establish that suitable engineering of external potentials serves as a powerful tool to generate supersolidity without requiring interaction tuning. This inspires unexplored pathways for unraveling nonequilibrium supersolidity, defect formation, and collective excitations in dipolar quantum gases. 
A next straightforward  direction is to investigate the defect formation caused by the barrier sweep in an initially dipolar superfluid state aiming to unveil emission of dark soliton structures similar to past observations in short-range interacting gases~\cite{Engels_obstacle}.  
Another interesting possibility is to explore whether similar supersolid creation can be achieved in higher dimensions where additional degrees-of-freedom are involved and distinct droplet lattices such as triangular or hexagonal configurations are formed depending on the atom number, external potential and interactions. 
Similarly, the study of defect generation in higher-dimensional dipolar superfluid states is also desirable for achieving nucleation of vortical patterns~\cite{Recati_vortex,casotti_observation_2024} especially in the superfluid-to-supersolid crossover.  
Finally, considering a rotating Gaussian barrier of an initial two-dimensional supersolid configuration for dynamically nucleating vortex lattices and possibly enter the turbulence regime~\cite{sabari_vortex_2024} is also a topic of interest.

\section*{Acknowledgments} 
S.I.M acknowledges support by the Army Research Office under Award number: W911NF-26-1-A043.  
E. L. B acknowledges  financial support from the OURE program of the Missouri University of Science and Technology. 
S.I.M and G.A.B are thankful to H. R. Sadeghpour, T. Bland, K. Mukherjee, and P. Giannakeas, for relevant discussions on the concept of supersolidity.

\appendix


\section{Role of three-body losses on the supersolid generation}\label{app:3body}

In the main text, we have demonstrated the dynamical nucleation of supersolid configurations by considering a potential barrier crossing a quasi-1D droplet lattice with different initial velocities. 
For this investigation, a closed system description was assumed based on the eGPE model of  Eq.~(\ref{Eq:MF_equation}). 
On the other hand, both droplet and supersolid states in relevant ultracold dipolar gas experiments undergo three-body losses due to their large densities~\cite{Maier_losses,Bottcher_losses,Chomaz2018}. 
The latter in turn may hinder the long-time observation of droplets and supersolids. 
As such, below we aim to explore the impact of such processes for our setup in order to showcase that supersolid formation is viable in the course of the evolution even in the presence of these mechanisms.

To simulate the corresponding nonequilibrium quantum dynamics we again employ the eGPE described by Eq.~(\ref{Eq:MF_equation}) but with the additional imaginary term contribution $-(i\hbar K_3 /2) \abs{\Psi(\boldsymbol{r},t)}^4 \Psi(\boldsymbol{r},t)$ which phenomenologically captures three-body loss mechanisms. Detailed discussions on loss processes in dipolar gas experiments and their competition with beyond mean-field effects can be found in Ref.~\cite{chomaz_dipolar_2022}. 
This imaginary term is characterized by the three-body recombination rate coefficient $K_3$~\cite{Bisset2015,Xi2016,Halder_control_2022}, which is taken here to be  $K_3=1.2\times10^{-40}~m^6/s$ as it was reported in the experiment of Ref.~\cite{ferrier2016observation}. 

\begin{figure}[t!]
\centering\includegraphics[width=1.0\linewidth]{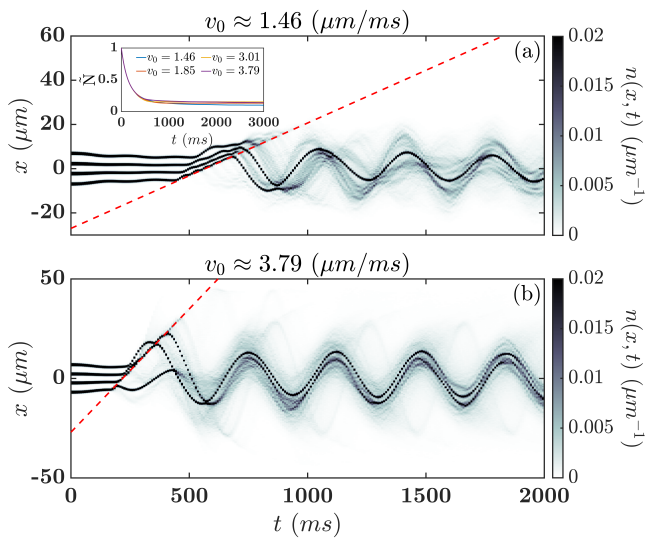}
\caption{Density dynamics of the dipolar gas under barrier sweeps with different velocities (see legends) in the presence of three-body recombination. 
Supersolid nucleation takes place after dragging the potential barrier in both cases. 
It is evident that the effect of three-body losses in the ensuing dynamical response is more prominent for smaller velocities; compare with Fig.~\ref{fig:Density_dynamics}(a), (d).  
Inset presents the normalized atom number, $\tilde{N}(t)=N(t)/N(0)$, during the time evolution for various velocities of the potential barrier (see legend). 
The system is initiated in the four crystal droplet array discussed in the main text and the remaining system parameters are the same as in Fig.~\ref{fig:Density_dynamics}.
}
\label{fig:3b_loss}
\end{figure}

The Dysprosium gas is prepared in its ground-state four crystal droplet with $a=86~a_0$, as in the main text. 
The emergent density evolution when the system is subjected to a barrier sweep with velocities $v_0=1.46~  \mu m/ms$ and $v_0=3.79~ \mu m/ms$ is visualized in Fig.~\ref{fig:3b_loss}(a) and (b) respectively. 
It is evident that in both cases the inclusion of three-body losses does not inhibit the formation of the supersolid structure. 
However, the resulting configurations structurally deviate for the closed system predictions, compare in particular Fig.~\ref{fig:Density_dynamics}(a) with Fig.~\ref{fig:3b_loss}(a) and Fig.~\ref{fig:Density_dynamics}(d) with Fig.~\ref{fig:3b_loss}(b). 
Indeed, even before the collision of the potential defect with the droplet lattice (stage \rm{I}) the presence of losses perturbs the crystals which start to drift toward the trap center, see Fig.~\ref{fig:3b_loss}. 
This is in contrast to the almost intact crystals in the absence of losses [Fig.~\ref{fig:Density_dynamics}(a), (d)].  
Notably, after dragging the potential barrier (stages {\rm II} and {\rm III}) the resulting number and peak amplitude of the crystals are less when losses are taken into account and the accompanying superfluid background is smaller. 
Moreover, for smaller velocities such as $v_0=1.46~ \mu m/ms$, the post-collision crystals of the supersolid undergo an oscillatory motion on top of the superfluid background. 
This response is apparently different from the one of the closed system [Fig.~\ref{fig:Density_dynamics}(a)], where the crystals perform relatively much smaller amplitude oscillations.

Arguably, more prominent differences occur for relatively smaller barrier velocities, e.g. $v_0=1.46~\mu m/ms$, since in these cases the potential defect hits the droplet lattice at later evolution times where a more significant atom fraction has been lost from the original sample. 
The normalized atom number, $\tilde{N}(t)=N(t)/N(0)$, is presented in the inset of Fig.~\ref{fig:3b_loss}(a). 
Independently of the barrier velocity there is roughly $20 \%$ number of atoms left after $t>1000~ ms$ which naturally suppresses both the number of crystals and the superfluid background. 
It also turns out that the barrier's velocity slightly impacts the atom number at times $t>900~ ms$ by means that smaller velocities facilitate a somewhat enhanced atom loss, e.g. $\tilde{N}(t>1500~ ms) = 0.11$ for $v_0 = 1.46~ \mu m/ms$ and $ \tilde{N}(t>1500~ ms) = 0.15$ when $v_0=3.01~\mu m/ms$.  
This is traced back to the fact that for smaller velocities the atom loss is somewhat  enhanced since the barrier collides later with the droplet structure, and thus the initial highly localized crystals facilitating atom decay remain un-perturbed during a longer time.

\begin{figure}[t!]
\centering\includegraphics[width=1\linewidth]{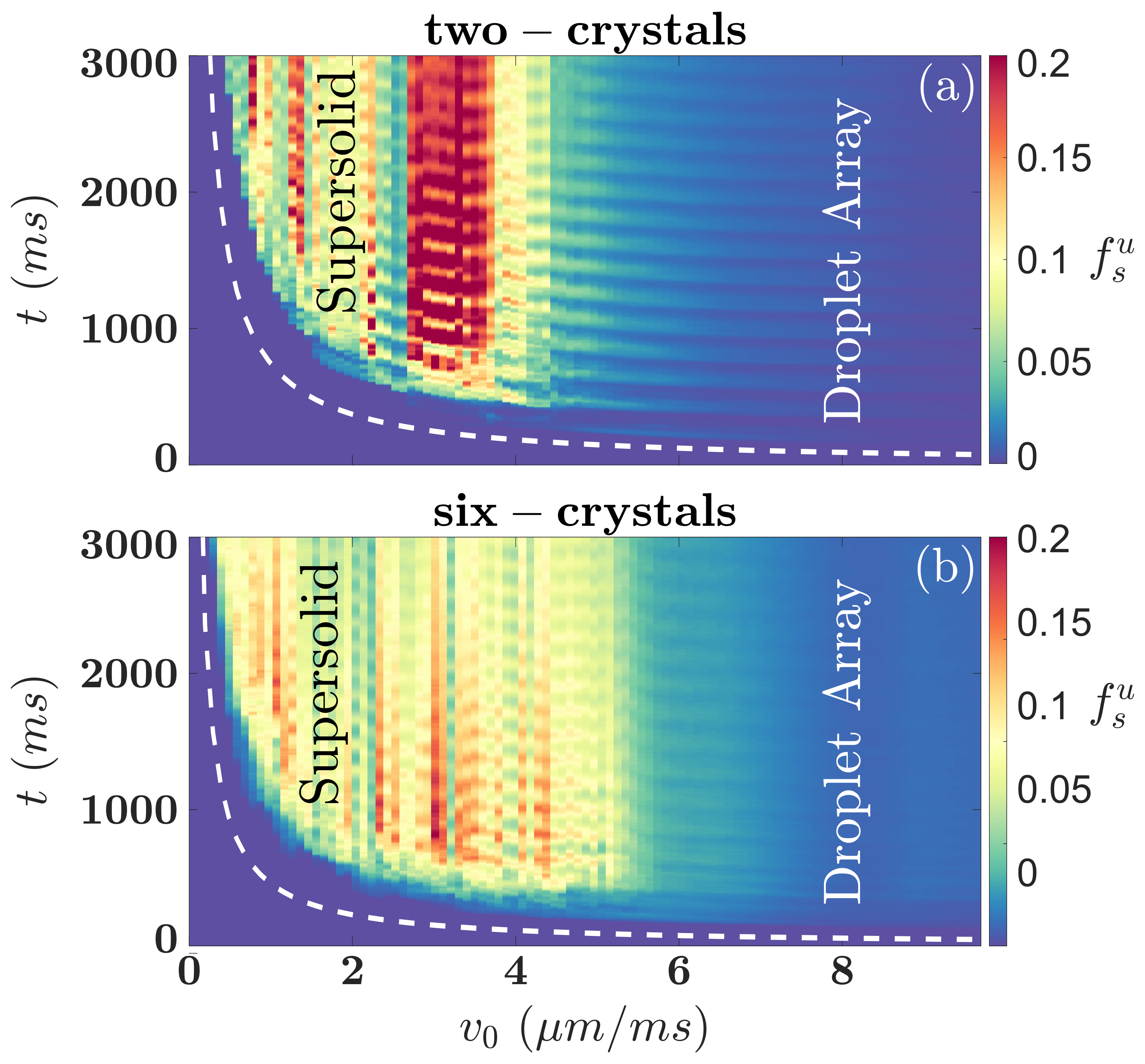}
\caption{Superfluid fraction pertaining to the dynamical response of an initial droplet array characterized by (a) two and (b) six droplet crystals.
Such configurations are realized by considering $N=25 \times 10^3$ and $N=25\times 10^4$ ${}^{164}$Dy atoms respectively at $a= 86 ~ a_0$. 
In both cases supersolids are dynamically generated stemming from the sweep of a Gaussian defect.
The white dashed lines correspond to the  
time $t^*$ of the first collision of the barrier with the nearest droplet peak. 
All other parameters are the same as in Fig.~\ref{fig:SF_fraction}. 
}
\label{fig:SF_fraction_v2}
\end{figure}

\section{Few to many droplet crystal response}\label{app:different_crystals}

To properly assess the dynamical nucleation of supersolids we need to also  address the impact of the initial state. In that regard, we consider $N=25\times 10^3$ and $N=25\times 10^4$ magnetic atoms (in contrast to $N=8\times 10^4$ presented in the main text) confined in the geometry described in Sec.~\ref{sec:theory}. The scattering length is set at  $a=86~a_0$, and the resulting ground states are droplet arrays.
In contrast to the four droplet peak arrays discussed in the main text, these particle numbers lead to two and six droplet crystals respectively.
Subsequently, these  ground-state configurations are perturbed utilizing  the barrier sweep protocol described by Eq.~\eqref{Eq:Gaussian_barrier}.
To facilitate direct comparison with the four crystal droplet response, we assume that the Gaussian potential possesses the same characteristics as in the main text.

To quantify the possibility of supersolid nucleation we invoke the superfluid fraction order parameter [see also Eq.~\eqref{Eq:Superfluid_fraction}], which is displayed in Fig.~\ref{fig:SF_fraction_v2} for both the two-crystal [Fig.~\ref{fig:SF_fraction_v2}(a)] and the six-crystal [Fig.~\ref{fig:SF_fraction_v2}(b)] configurations.
Initially, $f_s^u$ remains zero for all considered barrier velocities regardless of the particle numbers. Such a response holds only up to time instants $t^*$ (stage {\rm I}) preceding the barrier collision with the droplet array, denoted by the white dashed lines.

Similar to the scenario with four droplet peaks,  two distinct dynamical regimes of $f_s^u$ are discerned at longer evolution times belonging to consecutive  stages {\rm II} and {\rm III}. For large barrier velocities, $v_0 \gtrsim 6 ~ \mu m/ms$,  the droplet arrays remain almost unperturbed, as inferred by the zero superfluid fraction. 
The critical velocity beyond which the superfluid fraction remains zero has a weak dependence on the particle number.
In particular, the critical $v_0$ is slightly larger in the six-crystal initial state [c.f. Fig.~\ref{fig:SF_fraction_v2}(a) and (b)].
The six-crystal has a larger spatial extent and therefore an enhanced potential energy as compared to the smaller sized lattice.
As a result the binding energy, $\abs{\mu}$, of the $N=25 \times 10^4$ crystal is higher and therefore a larger barrier velocity is required to attain the diabatic regime, characterized by $v_0 \gg \sqrt{\abs{\mu}/m}$.

For smaller barrier velocities a clear enhancement of the superfluid fraction occurs regardless of the particle number.
On average, $f_s^u$ is slightly larger in the case of two crystals, e.g. at $v_0 = 4~ \mu m/ms$. 
Such a behavior is rooted in the partition of the defect's energy among many droplet peaks.
Every individual droplet peak becomes more excited in the case of smaller crystal arrays as compared to larger ones, since it can absorb a larger amount of energy. 
As a consequence, the avalanche mechanism triggered by the droplet collisions and leading to particle tunneling and emission described in Sec.~\ref{sec:collisions} is more efficient in the case of a smaller crystal.
Therefore, a more pronounced superfluid background can appear, i.e. a larger superfluid fraction.
In fact, the dynamical  configuration of the two-crystal case presented in Fig.~\ref{fig:SF_fraction_v2}(a), encompassing both the superfluid background and the crystals themselves, becomes significantly perturbed at long evolution times (not shown for brevity).

Similar to the four-droplet array addressed in the main text [Fig.~\ref{fig:SF_fraction}(a)], $f_s^u$ becomes resonantly enhanced at some specific barrier velocities.
The location of these resonances greatly depends on the particle number and hence the structure of the droplet array, alluding to the excitation of internal modes, which are distinct in every ground-state.
The persistence of supersolid generation for different atom numbers together with the dependence of the parametric regions of enhanced  superfluid background suggest the many-body character of the underlying mechanism.

\section{Numerical scheme}\label{app:simulations}

To obtain the ground-state and monitor the emergent nonequilibrium quantum dynamics of the 3D harmonically trapped dipolar gas under the influence of the moving Gaussian barrier, we numerically solve the eGPE described by Eq.~\eqref{Eq:MF_equation} in the main text.  
For convenience, the dimensionless form of the eGPE is utilized, where the 3D wave function is rescaled as $\Psi(\boldsymbol{r}, t) = \sqrt{N/l^3_x}\psi(\boldsymbol{r}', t')$, with the primed variables being  dimensionless. Also, the spatial and temporal scales are expressed with respect to the harmonic oscillator length $l_x=\sqrt{\hbar/m \omega_x}$ and the inverse trap frequency $\omega_x^{-1}$ of the elongated direction respectively. 
To numerically handle the dipolar interaction potential which diverges as ${\bf r}^{-3}$ at small interparticle distances, we employ the convolution theorem~\cite{arfken_mathematical_1972,Goral_ground_2002} 
\begin{align}
\int \text{d}^3\boldsymbol{r}' ~ U_\text{dd}(\boldsymbol{r} - \boldsymbol{r}') \abs{\Psi(\boldsymbol{r}',t)}^2 
= \mathcal{F}^{-1} \Big[ \mathcal{F}[U_\text{dd}] \cdot \mathcal{F}[\abs{\Psi}^2] \Big]. 
\end{align}
In this expression, $\mathcal{F}$ and $\mathcal{F}^{-1}$ represent the Fourier transform and its inverse respectively. 
Particularly, the Fourier transform of $U_\text{dd}$ is regular~\cite{Goral_ground_2002}, rendering the long-range interaction potential numerically well-defined and stable.

For the identification of the system's ground-state, especially in the droplet and supersolid regimes, it is crucial to use an appropriate wave function ansatz, based on symmetry considerations,  which in the quasi-1D geometry reads 
\begin{align}
\Psi(x,y,z) = \mathcal{A} \, e^{-\left( \frac{x^2}{2l_x^2} + \frac{y^2}{2l_y^2} + \frac{z^2}{2l_z^2} \right)} \sin^2\left(\frac{kx}{l_x}\right). 
\end{align}
Here, the parameter $k$ is varied in order to achieve the energetically lowest (ground) state configuration, while $\mathcal{A}$ is a normalization constant and $(l_y,l_z)=(\sqrt{\hbar/m \omega_y}, \sqrt{\hbar/m \omega_z})$ stand for the harmonic oscillator lengths across the transverse $y$- and $z$-directions.  
For our simulations, we employ a uniform 3D spatial grid characterized by  ($1024 \times 128 \times 128$) grid points to resolve the considered quasi-1D geometry with spatial discretization $\delta x = 0.0825~l_x$, $\delta y = 0.07~l_x$, $\delta z = 0.15~l_x$.

The time-evolution both in imaginary time, to identify the ground-state of the dipolar gas,  and in real time, for studying the ensuing nonequilibrium dynamics,  is addressed through the split-step Crank–Nicolson method~\cite{crank_practical_1947,antoine_computational_2013} deploying a time-step for the numerical integration $\delta t = 2 \times 10^{-4}/\omega_x$. These numerical parameters satisfy $(\omega_x \delta t)^2 < \delta x \delta y / l_x^2$~\cite{antoine_computational_2013} which ensures particle number and energy conservation in the course of the dynamics. 
Specifically, the normalization of the wave function at every step of the imaginary time propagation is maintained by applying $\psi(\boldsymbol{r}', t') \rightarrow \psi(\boldsymbol{r}',t')/||\psi(\boldsymbol{r}', t')||$. Numerical convergence of the order of $10^{-4}$ and $10^{-8}$ is attained for the ground-state wave function (at every grid point) and total energy respectively. 
Similarly, during the nonequilibrium dynamics (triggered by the moving Gaussian barrier) of the closed system the total particle number and total energy are preserved with accuracy $10^{-6}$ for the entire real time-evolution.

\twocolumngrid

\bibliographystyle{apsrev4-1}
\bibliography{Dipolar_Gases}

\end{document}